%% file: main.tex
\documentclass[sigconf, screen, authorversion]{acmart}
\input{macros}
\AtBeginDocument{%
  }

\copyrightyear{2025}
\acmYear{2025}
\setcopyright{rightsretained}
\acmConference[L@S '25]{Proceedings of the Twelfth ACM Conference on Learning @ Scale}{July 21--23, 2025}{Palermo, Italy}
\acmBooktitle{Proceedings of the Twelfth ACM Conference on Learning @ Scale (L@S '25), July 21--23, 2025, Palermo, Italy}
\acmDOI{10.1145/3698205.3733932}
\acmISBN{979-8-4007-1291-3/2025/07}

\begin{document}

\title{Bridging Gaps Between Student and Expert Evaluations of AI-Generated Programming Hints}



\author{Tung Phung}
\affiliation{%
  \institution{MPI-SWS}
  \city{Saarbrücken}
  \country{Germany}}
\email{mphung@mpi-sws.org}

\author{Mengyan Wu}
\affiliation{%
  \institution{University of Michigan}
  \city{Ann Arbor}
  \country{USA}}
\email{mengyanw@umich.edu}

\author{Heeryung Choi}
\affiliation{%
  \institution{University of Minnesota}
  \city{Twin Cities}
  \country{USA}}
\email{heeryung@umn.edu}

\author{Gustavo Soares}
\affiliation{%
  \institution{Microsoft}
  \city{Redmond}
  \country{USA}}
\email{gsoares@microsoft.com}

\author{Sumit Gulwani}
\affiliation{%
  \institution{Microsoft}
  \city{Redmond}
  \country{USA}}
\email{sumitg@microsoft.com}

\author{Adish Singla}
\affiliation{%
  \institution{MPI-SWS}
  \city{Saarbrücken}
  \country{Germany}}
\email{adishs@mpi-sws.org}

\author{Christopher Brooks}
\affiliation{%
  \institution{University of Michigan}
  \city{Ann Arbor}
  \country{USA}}
\email{brooksch@umich.edu}

\renewcommand{\shortauthors}{Tung Phung et al.}

\input{0_abstract}




\maketitle

\input{1_introduction}
\input{2_related_work}

\input{3_study_setup}
\input{4_results}

\input{5_next_step}
\input{6_conclusion}

\begin{acks}
Funded/Co-funded by the European Union (ERC, TOPS, 101039090). Views and opinions expressed are however those of the author(s) only and do not necessarily reflect those of the European Union or the European Research Council. Neither the European Union nor the granting authority can be held responsible for them.
\end{acks}

\clearpage

\bibliographystyle{ACM-Reference-Format}
\balance
\bibliography{main}



\end{document}

%% file: macros.tex
\usepackage[T1]{fontenc}
\usepackage{multirow}
\usepackage{subcaption}
\usepackage{xspace}
\usepackage{colortbl}
\usepackage[inline]{enumitem}
\usepackage{xcolor, soul}
\usepackage{arydshln}
\usepackage{setspace}
\usepackage{array}
\usepackage{makecell}
\usepackage{graphicx}
\usepackage{hyperref}
\usepackage{balance}

\newcommand{\textcode}[1]{{\fontfamily{cmtt}\selectfont #1}\xspace}

\definecolor{ExpHighlight}{rgb}{0.8, 0.1, 0.1}
\definecolor{mygreen}{rgb}{0,0.6,0}
\definecolor{mygray}{rgb}{0.5,0.5,0.5}
\definecolor{mymauve}{rgb}{0.58,0,0.82}
\definecolor{mynavy}{rgb}{0,0.133,0.302}
\definecolor{TextHighlight}{rgb}{1,1,0.6}
\definecolor{CodeGray}{rgb}{0.8,0.8,0.8}
\definecolor{CodeDarkGray}{rgb}{0.6,0.6,0.6}
\definecolor{OldPromptGrey}{rgb}{0.9,0.9,0.9}
\definecolor{NewPromptGreen}{rgb}{0.961,0.990,0.977}
\definecolor{InputOutput}{rgb}{0.764,0.345,0.009}
\definecolor{promptinputcolor}{rgb}{0.055,0.722,0.745}
\definecolor{promptheadercolor}{rgb}{0, 0, 0}

\newcommand{\promptinput}[1]{{\textcolor{promptinputcolor}{\textcode{#1}}}}

\usepackage{arydshln}  

\setlist{nolistsep,leftmargin=*}

\usepackage{listings}
\makeatletter
\let\old@lstKV@SwitchCases\lstKV@SwitchCases
\def\lstKV@SwitchCases#1#2#3{}
\makeatother
\usepackage{lstlinebgrd}
\makeatletter
\let\lstKV@SwitchCases\old@lstKV@SwitchCases

\lst@Key{numbers}{none}{%
    \def\lst@PlaceNumber{\lst@linebgrd}%
    \lstKV@SwitchCases{#1}%
    {none:\\%
     left:\def\lst@PlaceNumber{\llap{\normalfont
                \lst@numberstyle{\thelstnumber}\kern\lst@numbersep}\lst@linebgrd}\\%
     right:\def\lst@PlaceNumber{\rlap{\normalfont
                \kern\linewidth \kern\lst@numbersep
                \lst@numberstyle{\thelstnumber}}\lst@linebgrd}%
    }{\PackageError{Listings}{Numbers #1 unknown}\@ehc}}
\makeatother
\sethlcolor{TextHighlight}

\definecolor{vscode-pink}{RGB}{175,0,219} 
\definecolor{vscode-cyan}{RGB}{38,127,153} 
\definecolor{vscode-darkyellow}{RGB}{121,94,38} 
\definecolor{vscode-darkred}{RGB}{163,21,21} 
\definecolor{vscode-darkblue}{RGB}{0,15,128} 
\definecolor{vscode-green}{RGB}{0,153,0} 

%% file: 0_abstract.tex
\begin{abstract}
\looseness-1
Generative AI has the potential to enhance education by providing personalized feedback to students at scale. Recent work has proposed techniques to improve AI-generated programming hints and has evaluated their performance based on expert-designed rubrics or student ratings. However, it remains unclear how the rubrics used to design these techniques align with students' perceived helpfulness of hints. In this paper, we systematically study the mismatches in perceived hint quality from students' and experts' perspectives based on the deployment of AI-generated hints in a Python programming course. We analyze scenarios with discrepancies between student and expert evaluations, in particular, where experts rated a hint as high-quality while the student found it unhelpful. We identify key reasons for these discrepancies and classify them into categories, such as hints not accounting for the student's main concern or not considering previous help requests. Finally, we propose and discuss preliminary results on potential methods to bridge these gaps, first by extending the expert-designed quality rubric and then by adapting the hint generation process, e.g., incorporating the student's comments or history. These efforts contribute toward scalable, personalized, and pedagogically sound AI-assisted feedback systems, which are particularly important for high-enrollment educational settings.
\end{abstract}
%

%% file: 1_introduction.tex

\input{figs/illustration_prereflection_301/fig_main}

\section{Introduction}
\label{sec:introduction}

\looseness-1
Significant efforts have been dedicated to exploring the use of generative AI such as GPT-4~\cite{GPT4} for providing personalized programming feedback to students \cite{DBLP:conf/iticse/AzaizKS24,DBLP:conf/chi/KimmelGYGHOVWY24,DBLP:conf/sigcse/WoodrowMP24}. However, ensuring that automated feedback benefits learning is non-trivial as effective feedback should be both pedagogically sound from expert educators' perspectives and perceived as helpful by students~\cite{boud2012feedback,kunter2007expert,otaki2024generative}.
Motivated by this challenge, this study investigates the differences between students' and experts' evaluation of AI-generated hints, aiming to optimize hints to meet both students' and experts' quality criteria.

\looseness-1
Prior research has proposed techniques to enhance AI-generated hints' quality, often by using expert-designed scoring rubrics as benchmarks for optimization and evaluation~\cite{DBLP:conf/icer/PhungPCGKMSS22,DBLP:conf/lak/PhungPS0CGSS24,DBLP:conf/ace/RoestKJ24,neurips2023gaied_32_zamfirescu-pereira}. The use of rubric-based expert evaluation facilitates rapid optimization of hint-generation techniques while reducing the need for students’ evaluations, thereby avoiding possible negative impacts from exposing students to misleading information provided by premature techniques.
Some other studies have deployed AI-generated hints in classrooms and examined student evaluations~\cite{DBLP:conf/kolicalling/LiffitonSS023,DBLP:conf/sigcse/WangMP24,DBLP:journals/corr/abs-2406-05600}. 
Despite these efforts, there remain gaps in understanding the alignment between students' perceived helpfulness of hints and expert rubrics, as well as how techniques can be adjusted to address these discrepancies to enhance the overall quality of AI-generated hints. 

\looseness-1This paper systematically investigates student and expert ratings of AI-generated hints in a Python programming course with $74$ students.
Specifically, we focus on the following research questions:
\begin{itemize}[leftmargin=5pt,itemsep=1pt,topsep=3pt,label={}]
    \item \textbf{RQ1:} Are there differences in perceived hint quality from students' and experts' perspectives on AI-generated hints?
    \item \textbf{RQ2:} What are the main reasons for these discrepancies?
\end{itemize}
Based on the results, we further propose extensions to the rubric to better align with students' perceptions of hints and adjustments to the technique to improve hint quality and helpfulness.
\autoref{fig:illustration_prereflection_301} illustrates an example from our study.
Our findings form the foundation to advance the quality and helpfulness of AI-generated programming hints, thereby enhancing students' learning experiences.

%% file: figs/illustration_prereflection_301/fig_main.tex
\begin{figure*}
    \begin{minipage}{0.445\linewidth}
    {
        \begin{subfigure}{\linewidth}
        {
            \centering
            \begin{tabular}{|p|}
                \hline
                \multicolumn{1}{|c|}{\textbf{Problem description} (adjusted for brevity)} \\
                \multicolumn{1}{|p{\linewidth}|}{
                    {\small \input{figs/illustration_prereflection_301/content/task_very_short}}
                }\\
                \hline
            \end{tabular}
            \begin{tabular}{|p{\linewidth}|}
                \hline
                \multicolumn{1}{|c|}{\textbf{Student program}} \\
                \multicolumn{1}{|p{\linewidth}|}{
                    \hspace{2mm}
                    \scalebox{0.9}{
                        \renewcommand{\arraystretch}{1.6}
\begin{lstlisting}[basicstyle=\fontsize{8}{8.4}\ttfamily,basewidth=0.5em]
def (*@\textcolor{vscode-darkyellow}{load\_ticket\_data}@*)():
  (*@\textcolor{mygray}{<...import statements...>}@*)
  (*@\textcolor{vscode-cyan}{warnings}@*).(*@\textcolor{vscode-darkyellow}{filterwarnings}@*)('ignore')
  (*@\textcolor{vscode-darkblue}{df}@*) = (*@\textcolor{vscode-cyan}{pd}@*).(*@\textcolor{vscode-darkyellow}{read\_excel}@*)('assignments/assignment3/City-TicketViolation-jan2020.xls', (*@\textcolor{vscode-darkblue}{sheet\_name}@*)='name')
  (*@\textcolor{vscode-pink}{return}@*) (*@\textcolor{vscode-darkblue}{df}@*).(*@\textcolor{vscode-darkyellow}{head}@*)()
\end{lstlisting}
                     }  
                }
                \\
                \hline
            \end{tabular}	
            \begin{tabular}{|p{1\linewidth}|}
                \hline
                \multicolumn{1}{|c|}{\textbf{\cellcolor{OldPromptGrey}Hint provided by deployed technique}} \\
                \multicolumn{1}{|p{1\linewidth}|}
                {
                    {\small \input{figs/illustration_prereflection_301/content/old_hint}}
                }\\
                \hline
            \end{tabular}
            \vspace{-2mm}
            \label{fig:illustration_prereflection_301.scenario}
            \subcaption{Scenario}
        }
        \end{subfigure}
        \begin{subfigure}{\linewidth}
        {
            \vspace{0.5mm}
            \centering
             \begin{tabular}{|p{1\linewidth}|}
                \hline
                \multicolumn{1}{|p{1\linewidth}|}
                {            
                \centering
                \includegraphics[width=0.21\linewidth]{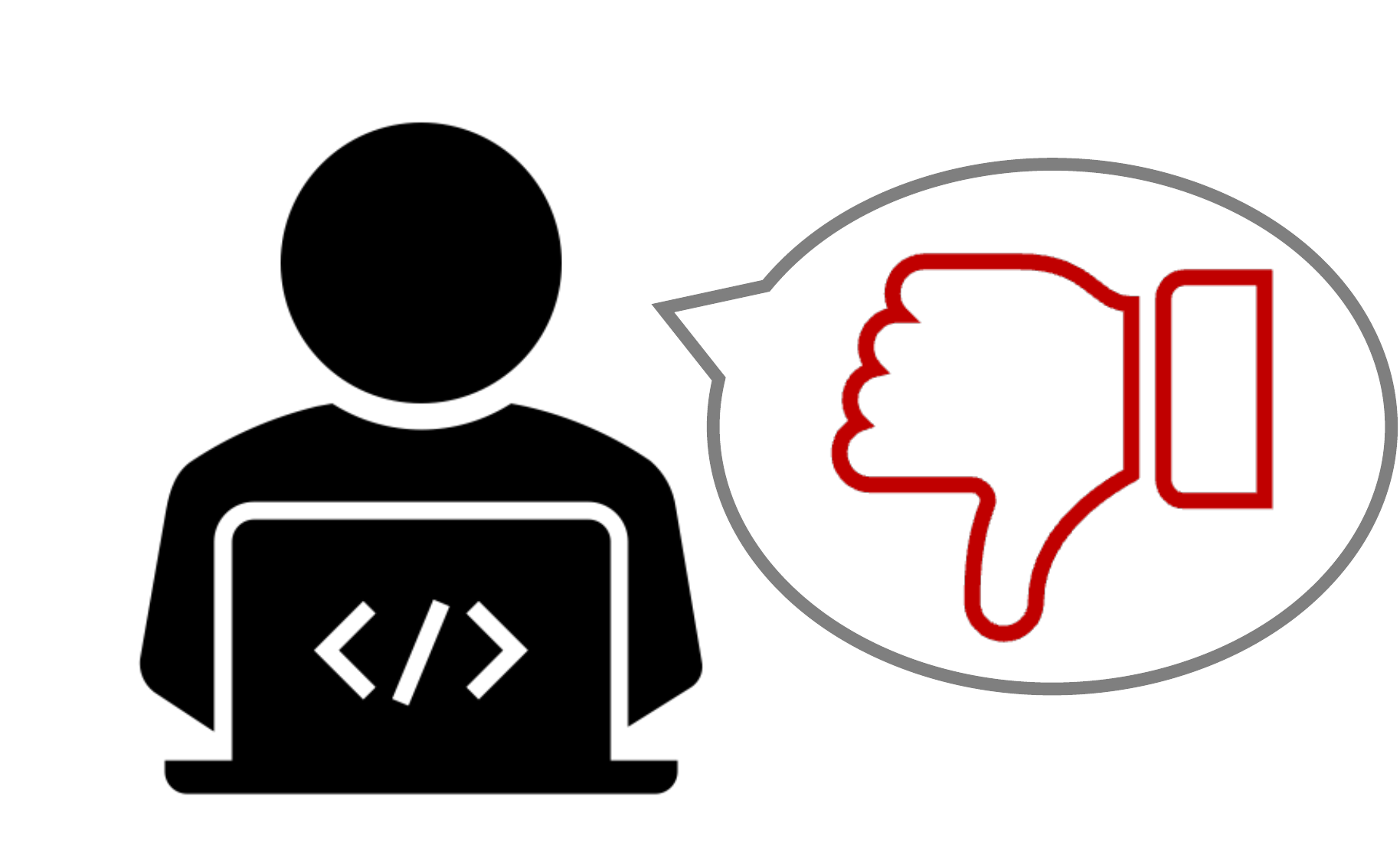}
                \quad \quad \quad
                \includegraphics[width=0.21\linewidth]{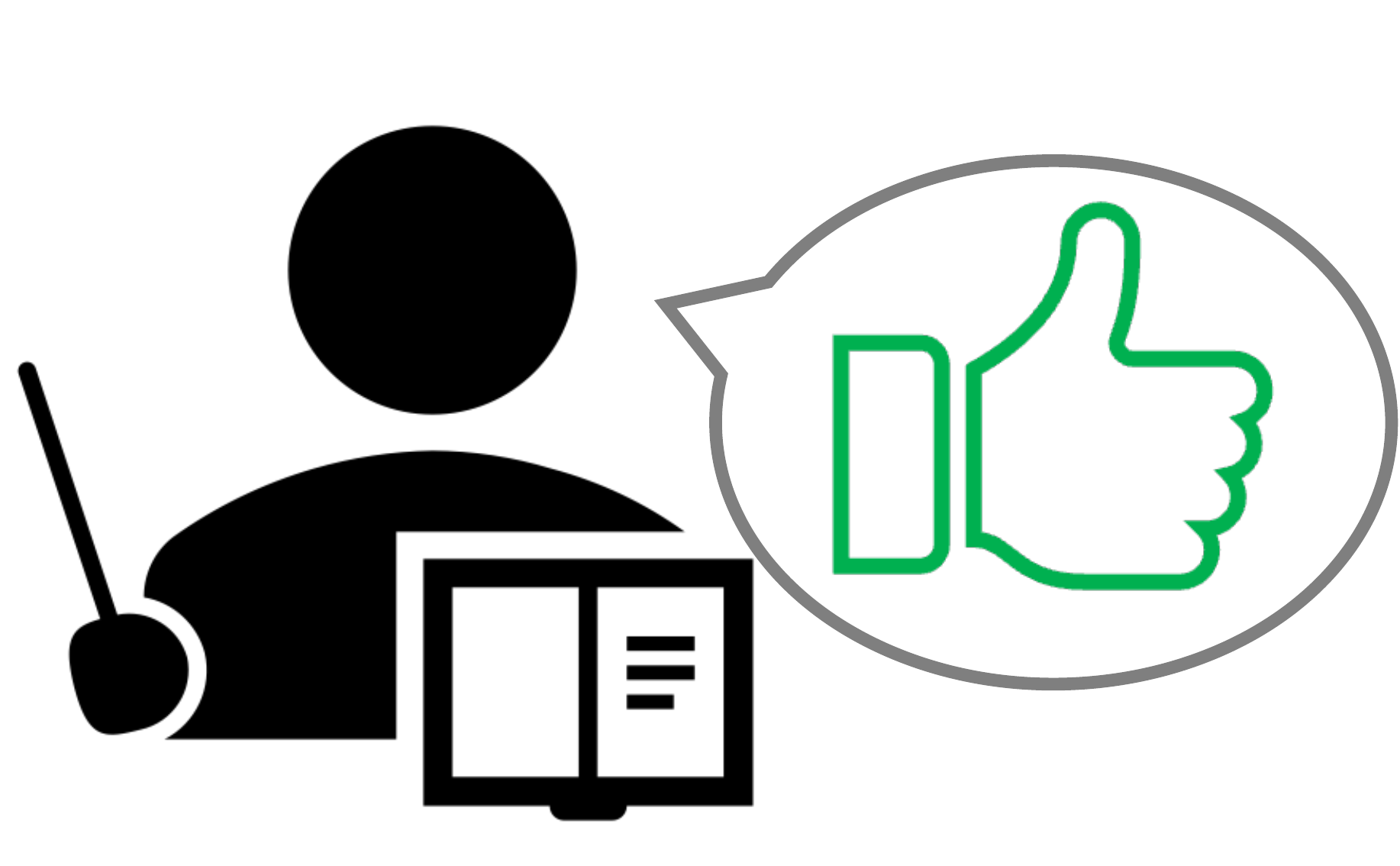}
                }
                \\
                \hline
            \end{tabular}
            \vspace{-2mm}
            \caption{Student and expert ratings of hint}				
            \label{fig:illustration_prereflection_301.ratings}
        }
        \end{subfigure}
    }
    \end{minipage}
    \hfill
    \begin{minipage}{0.52\linewidth}
    {
        \begin{subfigure}{0.5\linewidth}
        {
            \centering
            \begin{tabular}{|p{1\linewidth}|}
                \hline
                \multicolumn{1}{|p{1\linewidth}|}
                {                                   
                    {\small \input{figs/illustration_prereflection_301/content/student_reflection}}
                }\\
                \hline
            \end{tabular}
            \vspace{-2mm}
            \subcaption{Student's pre-hint thoughts}
        }
        \end{subfigure}
        \begin{subfigure}{0.435\linewidth}
        {
            \centering
             \begin{tabular}{|p{1\linewidth}|}
                \hline
                \multicolumn{1}{|p{1\linewidth}|}
                {                                   
                    {\small \input{figs/illustration_prereflection_301/content/mismatch_reason}}
                }\\
                \hline
            \end{tabular}
            \vspace{-2mm}
            \caption{Reason for rating mismatch}				
            \label{fig:illustration_prereflection_301.mismatch_reason}
        }
        \end{subfigure}
        \begin{subfigure}{\linewidth}
        {
            \vspace{1mm}
            \centering
             \begin{tabular}{|p{1\linewidth}|}
                \hline
                \multicolumn{1}{|c|}{\textbf{Prompt adjusted with Pre-hint Thoughts}} \\
                \multicolumn{1}{|p{1\linewidth}|}
                {                                   
                    {\small \input{figs/illustration_prereflection_301/content/prompt}}
                }\\
                \hline
            \end{tabular}		
            \begin{tabular}{|p{1\linewidth}|}
                \hline
                \multicolumn{1}{|c|}{\textbf{\cellcolor{NewPromptGreen}New hint after adjustment}} \\
                \multicolumn{1}{|p{1\linewidth}|}
                {                                 
                    {\small \input{figs/illustration_prereflection_301/content/prereflection_hint}}
                }\\
                \hline
            \end{tabular}
            \vspace{-2mm}
            \caption{Proposed adjustment}				
            \label{fig:illustration_prereflection_301.case_study}
        }
        \end{subfigure}
    }
    \end{minipage}
    \vspace{-8mm}
    \caption{
         \looseness-1Illustrative example. (a) shows the student's buggy program and provided hint. (b) shows the rating mismatch between student and experts. (c) shows student's thoughts while requesting hint, which helps explain the mismatch as shown in (d). (e) shows the adjusted prompt incorporating student's thoughts (\hl{changes highlighted in yellow}) and the resulting more helpful hint.
    }
    \label{fig:illustration_prereflection_301}
    \vspace{-3mm}
\end{figure*}

%% file: figs/illustration_prereflection_301/content/task_very_short.tex
Write code in function `load\_ticket\_data()' to load all Excel files in the current directory, concatenate the data, and return a single `DataFrame' object. The function should remove unnecessary headers and footers in each sheet. For this, you should scroll through every sheet in each file to determine the patterns for dropping irrelevant rows.

%% file: figs/illustration_prereflection_301/content/old_hint.tex
\cellcolor{OldPromptGrey}Consider the file path used in the buggy program. Is it correctly pointing to where the Excel files are actually located?

%% file: figs/illustration_prereflection_301/content/student_reflection.tex
Current issue is appropriate merge/concatenate of multiple sheets, since one file only has a single sheet.

%% file: figs/illustration_prereflection_301/content/mismatch_reason.tex
Hint does not address the student's concern as indicated by the pre-hint thoughts.

%% file: figs/illustration_prereflection_301/content/prompt.tex
I'm working on a Python programming problem. The current program below is not working well. Can you help by giving a hint?

\promptinput{\{problem\_description\}} 
\promptinput{\{current\_buggy\_output\}}
\promptinput{\{student\_program\}}
\promptinput{\{fixed\_program\}}
\promptinput{\hl{\{my\_thoughts\}}}

(1) Can you describe the bug(s) in this program and the required fixes? \hl{Consider my thoughts on possible issues.}
\newline
(2) Can you provide a hint about one bug in this program? \hl{Consider my thoughts on possible issues.} Do not give the answer or any code. If there's an obvious bug, direct me to the location of the bug. If there's a conceptual misunderstanding, offer me a conceptual refresher. Limit your response for the hint to a sentence or two at most. Be as socratic as possible, and be super friendly.

%% file: figs/illustration_prereflection_301/content/prereflection_hint.tex
\cellcolor{NewPromptGreen}Consider how the program handles the case where an Excel file contains only one sheet. Should it apply both the first sheet and last sheet rules to this single sheet?

%% file: 2_related_work.tex

\section{Related Work}

\looseness-1
\textbf{Rubrics for AI-generated content.} Rubrics are a well-established tool for teachers to evaluate the quality of students' work~\cite{jonsson2007use,reddy2010review}. Recently, rubrics are also used to assess AI-generated content~\cite{DBLP:conf/iticse/AzaizKS24,DBLP:conf/nips/KotalwarGS24,DBLP:conf/ace/RoestKJ24}. For instance, SPUR and RUBICON illustrated that rubrics effectively improve interpretability in human-generative AI interactions, underscoring the critical role of rubrics in domain-specific AI evaluation~\cite{DBLP:conf/aiware/BiyaniBRSG24,DBLP:journals/corr/abs-2403-12388}. This study contributes to this line of work by enriching expert-based rubrics for evaluating AI-generated hints.

\looseness-1\textbf{Techniques for AI-generated hints.} Early work has included the student's buggy program and the problem description in the prompt for generative AI to generate hints~\cite{DBLP:conf/edm/PhungCGKMSS23,DBLP:conf/ace/RoestKJ24}. To improve hint quality, later work has incorporated symbolic information such as failing test cases and fixed programs in the prompts, showing substantial improvement~\cite{DBLP:conf/lak/PhungPS0CGSS24}. 
In this study, we employ a hint-generation technique that incorporates symbolic information, presented through a button-based interface for requesting hints.

\textbf{Student and expert evaluations of AI-generated hints.}
\looseness-1While students may not always prefer the most instructionally beneficial options, they are generally better at assessing the cognitive demands of a task~\cite{kirschner2013learners,kunter2007expert}. In contrast, although experts can recommend better paths for learning gains, their perceptions of task difficulty or intervention quality can differ from those of students~\cite{beishuizen2001students,kunter2007expert}. Despite such discrepancies, many previous studies evaluated AI-generated hints using ratings from either students or experts, which could result in incomplete and biased evaluations~\cite{DBLP:conf/kolicalling/LiffitonSS023,DBLP:journals/corr/abs-2307-00150,DBLP:conf/lak/PhungPS0CGSS24}.
This work combines and investigates evaluations from both students and experts to provide a more holistic assessment of AI-generated hints.

%% file: 3_study_setup.tex

\section{Study and Analysis Setup}
\label{sec:setup}
\looseness-1
\textbf{Study context.}
This study was conducted in a Python-based introductory data science programming course as part of an online Master's program at the University of Michigan. This four-week course featured weekly assignments, each was provided as a Jupyter notebook containing three to four programming questions. The questions covered data manipulation skills using regular expressions and the \textit{pandas} library. 
Students were informed that requesting hints is voluntary and anonymous, there is no additional incentive or penalty, and that hints are AI-generated and might not always be correct. This study was reviewed by the local Institutional Review Board (IRB) and approved as being exempt from oversight. 

\looseness-1
\textbf{Deployed technique for AI-generated hints.}\footnote{\url{https://github.com/machine-teaching-group/las2025-bridging-gaps-ai-hints}} 
We adapted the technique from previous studies that showed good performance in data science education~\cite{DBLP:conf/lak/PhungPS0CGSS24,neurips2023gaied_32_zamfirescu-pereira}.
The deployed technique, using GPT-4, operates in two stages: First, given the student's buggy program, it extracts symbolic information including the buggy output (by running the buggy program) and a fixed program (by requesting the generative AI model). 
Second, it uses all this information to request the AI model to generate an explanation (for leveraging the Chain-of-Thought effects~\cite{DBLP:conf/nips/Wei0SBIXCLZ22}) and subsequently a Socratic-style hint for one bug in the student's program. For pedagogical reasons, we provided only the hint to the student. The hint-generation prompt is the one shown in \autoref{fig:illustration_prereflection_301.case_study} (top) without the yellow parts.

\looseness-1
\textbf{Student interaction and ratings.}
\autoref{fig:workflows.data_collection} illustrates student interaction with our system. We developed a JupyterLab extension as the interacting interface with students and a backend to run the technique. To be able to request hints, students needed to click an activation button and give consent for data to be recorded anonymously. Out of the total $74$ students, $34$ ($46\%$) activated and requested hints. After activating, students would see a \textit{Hint} button next to each question in any assignment to request hints. To prevent over-reliance on the system, students are limited to three hints per question. After receiving a hint, students were asked to rate it as either \textit{Helpful} or \textit{Unhelpful}. We note that students were encouraged to identify and reflect on their possible issues before and after a hint as an optional activity; however, their responses (i.e., pre-hint and post-hint thoughts on issues) were \textit{not} used to generate hints.

\input{figs/workflows/data_collection/fig_main}

\looseness-1
\textbf{Rubric and expert ratings.}
After the course, two experts discussed and rated each hint together. Both of them had extensive experience in teaching Python. One previously taught earlier runs of this course. The experts rated the hints following a rubric from literature~\cite{DBLP:conf/nips/KotalwarGS24,DBLP:conf/lak/PhungPS0CGSS24}, which comprises four binary attributes: 
\textit{Correct} (providing correct information regarding issue(s) in the program); 
\textit{Informative} (providing useful information to help resolve the issue(s)); 
\textit{Conceal} (providing not too-detailed information, so the student would also have to reason about implementing the fixes);
\textit{Comprehensible} (being readable, easy to understand, and containing no redundant information). 
A hint was rated as high-quality if all these attributes were satisfied, and rated as low-quality otherwise.

\looseness-1
\textbf{Analysis setup.}
\autoref{fig:workflows.data_analysis} illustrates our data analysis workflow. 
For RQ1, we compare experts' and students' ratings of hints using a $2\times2$ contingency table.
For RQ2, we identify cases where students' and experts' ratings did not match. Specifically, we focus on cases where experts rated hints as high-quality but students found them unhelpful, and we categorize reasons for these discrepancies.

\input{figs/workflows/data_analysis/fig_main}

%% file: figs/workflows/data_collection/fig_main.tex
\begin{figure}[h]
    \centering
    \scalebox{1}{
        \includegraphics[height=0.11\paperheight]{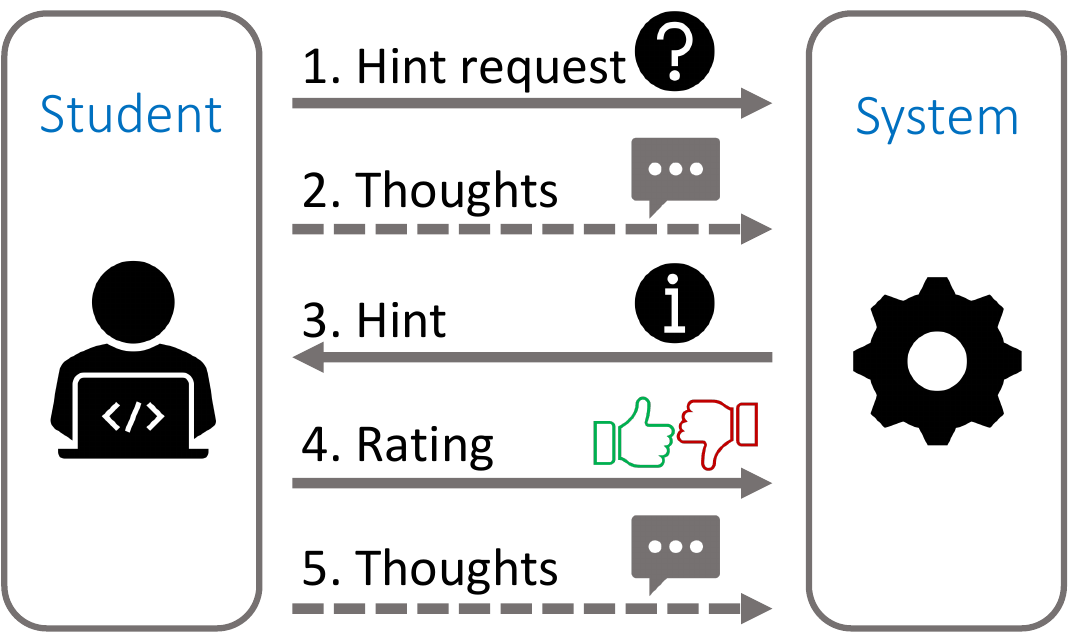}
    }
    \vspace{-3.5mm}
    \caption{
        Interaction between students and our system. It is optional for students to give thoughts on their current issues.
    }
    \label{fig:workflows.data_collection}
    \vspace{-2mm}
\end{figure}

%% file: figs/workflows/data_analysis/fig_main.tex
\begin{figure}[h]
    \vspace{-2mm}
    \centering
    \scalebox{1}{
        \includegraphics[height=0.11\paperheight]{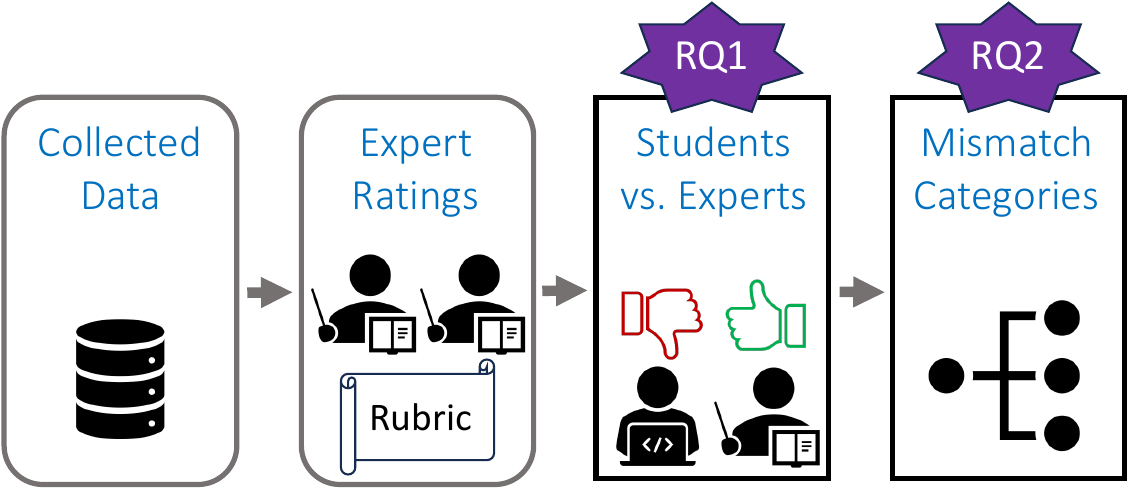}
    }
    \vspace{-3mm}
    \caption{
        Data analysis workflow. After experts rate the hints, the two research questions are addressed in sequence.
    }
    \label{fig:workflows.data_analysis}
    \vspace{-4mm}
\end{figure}

%% file: 4_results.tex

\vspace{-1mm}
\section{Results for Research Questions}
\label{sec:results}


\textbf{RQ1: Students' vs. experts' ratings.}
Throughout the course, $34$ students received $158$ hints.
They rated $139$ of these hints, which were then compared with experts' ratings.
As summarized in \autoref{table:result_ratings_students_experts}, agreement between students and experts was observed in 91 cases ($65.5\%$) and disagreement in 48 cases ($34.5\%$). The difference between students' and experts' ratings is significant w.r.t a $\chi^2$ test~\cite{f3e1a971-f77a-3568-83d6-8c0b7f53e9e5} ($p\leq0.001$). Specifically, 37 out of 48 cases of mismatch ($77\%$) occurred when experts rated a hint as high-quality while students found it unhelpful. 
This finding suggests that rubric-based expert evaluations may be overly optimistic, necessitating further diagnosis and improvement.
\footnote{We also investigated $11$ cases where experts rated hints as low-quality but students found them helpful. Most of these hints are partially incorrect, thereby violating the rubric, though students potentially learned from the correct parts to fix their bugs and thus rated them helpful. In our analysis, we primarily focus on the cases where students rated the hints as unhelpful and seek to provide more helpful hints.}

\input{figs/results/ratings_students_experts/fig_main}

\looseness-1
\textbf{RQ2: Mismatch categories.}
To understand the causes of these discrepancies, we focus our analysis on the $37$ cases where experts overestimated hint quality.
Our categorization of mismatch is guided by the general characteristics of feedback~\cite{boud2012feedback,ferguson2011student,snyder2008teaching,voelkel2020students} and informed by the patterns observed in the collected data.
As summarized in \autoref{table:result_ratings_categories}, the mismatches with deducible reasons are categorized as follows:
\begin{itemize}[label=$\bullet$]
    \item \looseness-1\textit{Mismatch in pedagogical objectives.} Hints are correct and balanced in offering useful information without revealing the solution, yet students appeared to desire more detailed assistance.
    
    \item \looseness-1\textit{Student's concern is not addressed.} Hints are often personalized based on the student's code, but the code alone does not always capture their main concern (see \autoref{fig:illustration_prereflection_301}).

    \item \looseness-1\textit{Student's trajectory is ignored.} When a student requests multiple hints for the same question, later hints sometimes give no additional information regarding previous ones.

    \item \looseness-1\textit{Struggle in solving approach is ignored.} When a student's program is empty or is on a completely wrong track, hints addressing a low-level bug are not adequate and relevant.

    \item \textit{Student's progress is not acknowledged.} When students are on the right track, hints only stating issues related to incomplete code do not provide the necessary guidance and acknowledgment. 
 
\end{itemize}
\looseness-1For the remaining $11$ cases, the reasons for students finding hints unhelpful are not evident from the available data.

\input{figs/results/mismatch_categories/fig_main}


%% file: figs/results/ratings_students_experts/fig_main.tex
\begin{table}[h]
    \centering
    \vspace{-1mm}
    \caption{
        RQ1: Comparison of students’ and experts’ ratings.
    }
    \vspace{-3mm}
    \scalebox{0.9}{
    \setlength\tabcolsep{3pt}
    \renewcommand{\arraystretch}{1.18}
    \input{figs/results/ratings_students_experts/table}

    }
    \label{table:result_ratings_students_experts}
    \vspace{-1mm}
\end{table}

%% file: figs/results/ratings_students_experts/table.tex
\begin{tabular}{cc|cc}
    \hline
    \multicolumn{2}{c|}{\multirow{3}{*}{}} & \multicolumn{2}{c}{\textbf{Experts' ratings}} \\
     & & \multicolumn{1}{c}{High-quality}& \multicolumn{1}{c}{Low-quality}\\
     & & \includegraphics[width=0.075\linewidth]{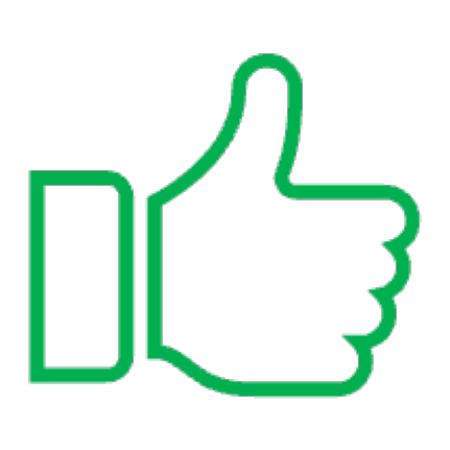} & \includegraphics[width=0.075\linewidth]{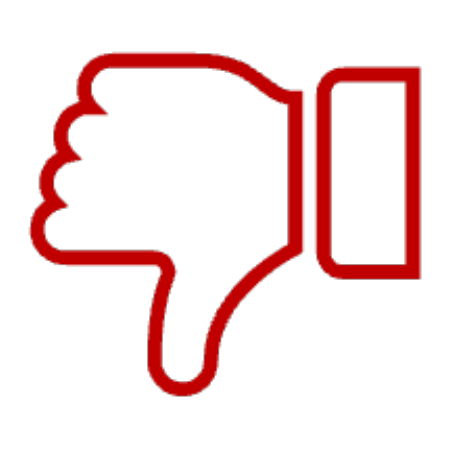} \\
    \hline
    \multicolumn{1}{c}{\multirow{2}{*}{\textbf{Students' ratings}}} & \multicolumn{1}{r|}{Helpful \includegraphics[width=0.075\linewidth,trim={0 7mm 0 0}]{figs/misc/helpful.pdf}} & 68 (48.9\%) & 11 \phantom{0}(7.9\%)\\
    & \multicolumn{1}{r|}{Unhelpful \includegraphics[width=0.075\linewidth,trim={0 7mm 0 0},clip]{figs/misc/unhelpful.pdf}} & 37 (26.6\%) & 23 (16.5\%)\\
    \hline
\end{tabular}

%% file: figs/results/mismatch_categories/fig_main.tex
\begin{table}[h]
    \centering
    \vspace{-1.5mm}
    \caption{
        RQ2: Categorization of cases where experts rated the hints as high-quality but students found them unhelpful.
    }
    \vspace{-3.5mm}
    \scalebox{1}{
        \setlength\tabcolsep{10pt}
        \renewcommand{\arraystretch}{1}
        \input{figs/results/mismatch_categories/table}
    }
    \label{table:result_ratings_categories}
    \vspace{-4mm}
\end{table}

%% file: figs/results/mismatch_categories/table.tex

\begin{tabular}{lc}
    \hline
    \multicolumn{1}{c}{\textbf{Mismatch category}} & \textbf{Cases} \\
    \hline
    Reason known & \fontsize{8}{7}\selectfont 26 (70.3\%)\\
    \quad Mismatch in pedagogical objectives & \fontsize{8}{7}\selectfont 10 (27.0\%)\\
    \quad Student’s concern is not addressed & \fontsize{8}{7}\selectfont \phantom{0}8 (21.6\%)\\
    \quad Student's trajectory is ignored & \fontsize{8}{7}\selectfont \phantom{0}4 (10.8\%)\\
    \quad Struggle in solving approach is ignored & \fontsize{8}{7}\selectfont \phantom{0}2 \phantom{0}(5.4\%)\\
    \quad Student's progress is not acknowledged & \fontsize{8}{7}\selectfont \phantom{0}2 \phantom{0}(5.4\%)\\
    \hline
    \multicolumn{1}{l}{Reason unknown} & \fontsize{8}{7}\selectfont 11 (29.7\%)\\
    \hline
\end{tabular}

%% file: 5_next_step.tex

\input{figs/results/revised_rubric_and_prompt_adjustments/fig_main}

\section{Ideas for Improvement and Initial Findings}
\label{sec:ideas}

Our ongoing next steps involve developing methods to bridge gaps and enhancing the overall quality of AI-generated programming hints. Since resolving pedagogical conflicts is not the main consideration of this work, we focus on the other four mismatch categories with known reasons as described in Section~\ref{sec:results}. Based on the results above, we propose an extension to the current rubric to better align student and expert ratings. Further, we propose adjustments to improve the hint-generation process for higher-quality hints.

More concretely, we extend the rubric described in Section~\ref{sec:setup} (including attributes Correct, Informative, Conceal, and Comprehensible) by augmenting it with four additional quality attributes (\autoref{table:revised_rubric}). Preliminary results show that experts following this \emph{extended rubric} rate all $16$ hints in the four mismatch categories of interest as low-quality, suggesting a higher alignment with students in cases the current rubric was insufficient. However, the extended rubric might also render more hints to be recognized as low-quality, necessitating enhancements in hint-generation techniques.

We based on the new rubric to propose potential adjustments to the hint-generation process (\autoref{table:prompt_adjustments}). \textit{Pre-hint Thoughts} and \textit{Post-hint Thoughts} adjustments convey the students' main concerns to be addressed, exemplifying a focused form of back-and-forth communications between humans and AI~\cite{DBLP:journals/corr/abs-2402-06229,DBLP:journals/corr/abs-2307-01644,DBLP:conf/kolicalling/LiffitonSS023,DBLP:journals/corr/abs-2403-19154}, directing AI towards discussing specific issues~\cite{lohr2025you}.
\textit{Trajectory} adjustment incorporates the history of student interaction and additional context for how new hints should be adapted to~\cite{neurips2023gaied_32_zamfirescu-pereira,DBLP:journals/corr/abs-2401-05319,DBLP:conf/icer/Singla22}.
\textit{Overarching Bug} and \textit{Solving Plan} adjustments represent a form of expert knowledge-based instruction in the prompt, directing the model towards a suitable type of feedback~\cite{remadi2024prompt,DBLP:conf/nips/BrownMRSKDNSSAA20}.
In our initial findings, these adjustments address the $16$ mismatch cases above, generating hints meeting the \emph{extended} rubric, suggesting to be more helpful to students (see \autoref{fig:illustration_prereflection_301} for an illustration of Pre-hint Thoughts adjustment; more examples are provided in our public repository).

%% file: figs/results/revised_rubric_and_prompt_adjustments/fig_main.tex
\begin{table*}[t!]
    \centering
    \caption{
        Our proposals to better align students' and experts' ratings of hints and to improve hint quality.
    }
    \vspace{-4mm}
    \begin{minipage}{\linewidth}
        \begin{subfigure}{\linewidth}
            \subcaption{Our proposed rating attributes to augment the rubric}
            \vspace{-1.5mm}
            \scalebox{0.82}{
                \setlength\tabcolsep{5.3pt}
                \renewcommand{\arraystretch}{1.1}
                \input{figs/results/revised_rubric_and_prompt_adjustments/revised_rubric_table}

            }
            \label{table:revised_rubric}
        \end{subfigure}
        \vspace{0.75mm}
    \end{minipage}
    \begin{minipage}{\linewidth}
        \begin{subfigure}{\linewidth}
            \subcaption{Our proposed adjustments to generate better hints}
            \vspace{-1.5mm}
            \scalebox{0.82}{
                \setlength\tabcolsep{5pt}
                \renewcommand{\arraystretch}{1.05}
                \input{figs/results/revised_rubric_and_prompt_adjustments/prompt_adjustments_table}
            }
            \label{table:prompt_adjustments}
        \end{subfigure}
    \end{minipage}
    \vspace{-2mm}
    \label{table:revised_rubric_and_prompt_adjustments}
\end{table*}

%% file: figs/results/revised_rubric_and_prompt_adjustments/revised_rubric_table.tex

\begin{tabular}{c|c}
    \hline
    \textbf{New Attributes} & \textbf{Description} \\
    \hline
    \multicolumn{1}{p{0.154\linewidth}|}{Accounting\newline student’s concern} 
    & \multicolumn{1}{p{0.999\linewidth}}{Focusing on the student's main concern. For instance, when a student has trouble understanding the problem description, the hint should clarify confusion about the problem rather than help solve a bug.} \\
    \hline
    \multicolumn{1}{p{0.154\linewidth}|}{Informative\newline given history} 
    & \multicolumn{1}{p{0.999\linewidth}}{Containing new valuable information that was not available in previous hints, e.g., by describing another bug, explaining a new approach to detect or fix the bug, or giving more information about a previously mentioned bug.} \\
    \hline
    \multicolumn{1}{p{0.154\linewidth}|}{Tackling\newline overarching bug}
    & \multicolumn{1}{p{0.999\linewidth}}{When the hint is about solving a bug, it should address the most critical bug. For instance, when a student's program contains both a conceptual bug and a mistyping, the conceptual bug should typically be prioritized.} \\
    \hline
    \multicolumn{1}{p{0.154\linewidth}|}{Guiding} 
    & \multicolumn{1}{p{0.999\linewidth}}{\looseness-1Conveying guidance relevant to the student’s current progress. For instance, when a student is far from solving the problem, hints may provide refreshers or offer solving plans; when a student is close to solving, hints can focus on bugs.} \\
    \hline
\end{tabular}

%% file: figs/results/revised_rubric_and_prompt_adjustments/prompt_adjustments_table.tex
\begin{tabular}{c|c}
    \hline
    \textbf{Adjustment} & \textbf{Description} \\
    \hline
    \multicolumn{1}{p{0.155\linewidth}|}{Pre-hint Thoughts} & \multicolumn{1}{p{1.0\linewidth}}{Asking the student for their thoughts on their current issues and using it in the prompt for hint generation.} \\
    \multicolumn{1}{p{0.155\linewidth}|}{Post-hint Thoughts} & \multicolumn{1}{p{1.0\linewidth}}{If the provided hint was not helpful, asking the student for their thoughts and using it in the prompt for a follow-up hint.} \\
    \multicolumn{1}{p{0.155\linewidth}|}{Trajectory} & \multicolumn{1}{p{1.0\linewidth}}{Incorporating the student’s trajectory (i.e., information from previous requests) into the prompt for hint generation.} \\
    \multicolumn{1}{p{0.155\linewidth}|}{Overarching Bug} & \multicolumn{1}{p{1.0\linewidth}}{Instructing the AI model to identify the overarching issue to focus the hint on.} \\
    \multicolumn{1}{p{0.155\linewidth}|}{Solving Plan} & \multicolumn{1}{p{1.0\linewidth}}{When a student is far from solving, requesting the model for a problem-solving plan w.r.t the student’s current progress.} \\
    \hline
\end{tabular}

%% file: 6_conclusion.tex
\section{Conclusion, Limitations, and Future Work}

This paper presents our in-progress study of the discrepancies between students' and experts' perspectives on the quality of AI-generated programming hints. Our analysis led to a systematic categorization of these discrepancies, informing the development of an extended rubric and proposals for adjustments to enhance AI-generated hints. The categorization of discrepancies provides a deeper understanding of students' struggles when interacting with AI-generated hints. The proposed rubric demonstrated a high alignment with students' ratings in cases where the original rubric falls short, facilitating offline development of hint-generation techniques without adversely impacting students. Lastly, our proposed adjustments to the hint-generation process showed potential for improving the quality and helpfulness of automated programming hints. These contributions support the broader goal of enhancing student learning at scale by optimizing automated hints for high-quality and personalized help--especially important in high-enrollment courses where giving manual teachers' feedback is infeasible.

We acknowledge some limitations and outline directions for future work. 
First, the current study is limited by its small sample size (34 students requested hints) from a single Python programming course, which constrains the generalizability of our findings. In the future, a more comprehensive understanding of student and expert rating discrepancies can be obtained from conducting experiments across diverse courses with larger and more varied student populations, thereby capturing a broader range of perspectives and enriching our insights.
Second, while our extended rubric better aligns with students in cases where the original rubric fails, it is important to examine its impacts on already-aligned cases.
Third, although our proposed technique adjustments have shown promise in enhancing hint quality, a mechanism should be developed to automatically integrate relevant adjustments to tailor to students' hint requests.
Further, it is crucial to thoroughly validate the effectiveness of these adjustments on hint quality and student learning outcomes through classroom deployments.